\documentclass [] {aa} 

\usepackage{txfonts,graphicx,lscape}

\begin{document}

\title{Abundance ratios of volatile vs.\ refractory elements in planet-harbouring stars: hints of pollution?
\thanks{Based on data from the {\footnotesize FEROS} spectrograph at the 2.2-m ESO/MPI 
    telescope (observing run ID 074.C-0135), at the La Silla Observatory, ESO (Chile), and the UVES 
    spectrograph at VLT/UT2 Kueyen telescope (observing run ID 074.C-0134), at the Paranal Observatory, 
    ESO (Chile), and on observations made with the SARG spectrograph at 
    3.5-m TNG, operated by the Fundaci\'on Galileo Galilei of the INAF, 
    and with the UES spectrograph at the 4-m William Hershel Telescope 
    (WHT), operated by the Isaac Newton Group, both at the Spanish 
    Observatorio del Roque de los Muchachos 
    of the Instituto de Astrofisica de Canarias.}}

\author{A.~Ecuvillon\inst{1}, G.~Israelian\inst{1}, N. C.~ Santos\inst{2,3},  M.~Mayor\inst{3} \and G.~Gilli\inst{1,4}}

\offprints{\email{aecuvill@ll.iac.es}}

\institute{Instituto de Astrof\'{\i}sica de Canarias, E-38200 La Laguna, Tenerife, Spain \and Centro de 
Astronomia e Astrofisica de Universidade de Lisboa, Observatorio Astronomico de Lisboa, Tapada de Ajuda, 1349-018 Lisboa, 
Portugal \and Observatoire de Gen\`eve, 51 ch.  des  Maillettes, CH--1290 Sauverny, Switzerland \and Dipartamento di Astronomia, Universit\'a di Padova, vicolo dell'Osservatorio 2, 35122 Padova, Italy
}
\date{Received 17 Nov 2005 / Accepted 02 Dec 2005} 

\titlerunning{Abundance ratios of volatile vs.\ refractory elements} 
\authorrunning{A. Ecuvillon et al.}

\abstract{We present the [$X$/H] trends as function of the elemental condensation temperature T$_C$ in 88 planet host stars and in a volume-limited comparison sample of 33 dwarfs without detected planetary companions. We gathered homogeneous abundance results for many volatile and refractory elements spanning a wide range of T$_C$, from a few dozens to several hundreds kelvin. We investigate possible anomalous trends of planet hosts with respect to comparison sample stars in order to detect evidence of possible pollution events. No significant differences are found in the behaviour of stars with and without planets. This result is in agreement with a ``primordial'' origin of the metal excess in planet host stars. However, a subgroup of 5 planet host and 1 comparison sample stars stands out for having particularly high [$X$/H] vs.\ T$_C$ slopes.
	  }
\maketitle

\begin{figure*}
\centering 
\includegraphics[width=6.7cm]{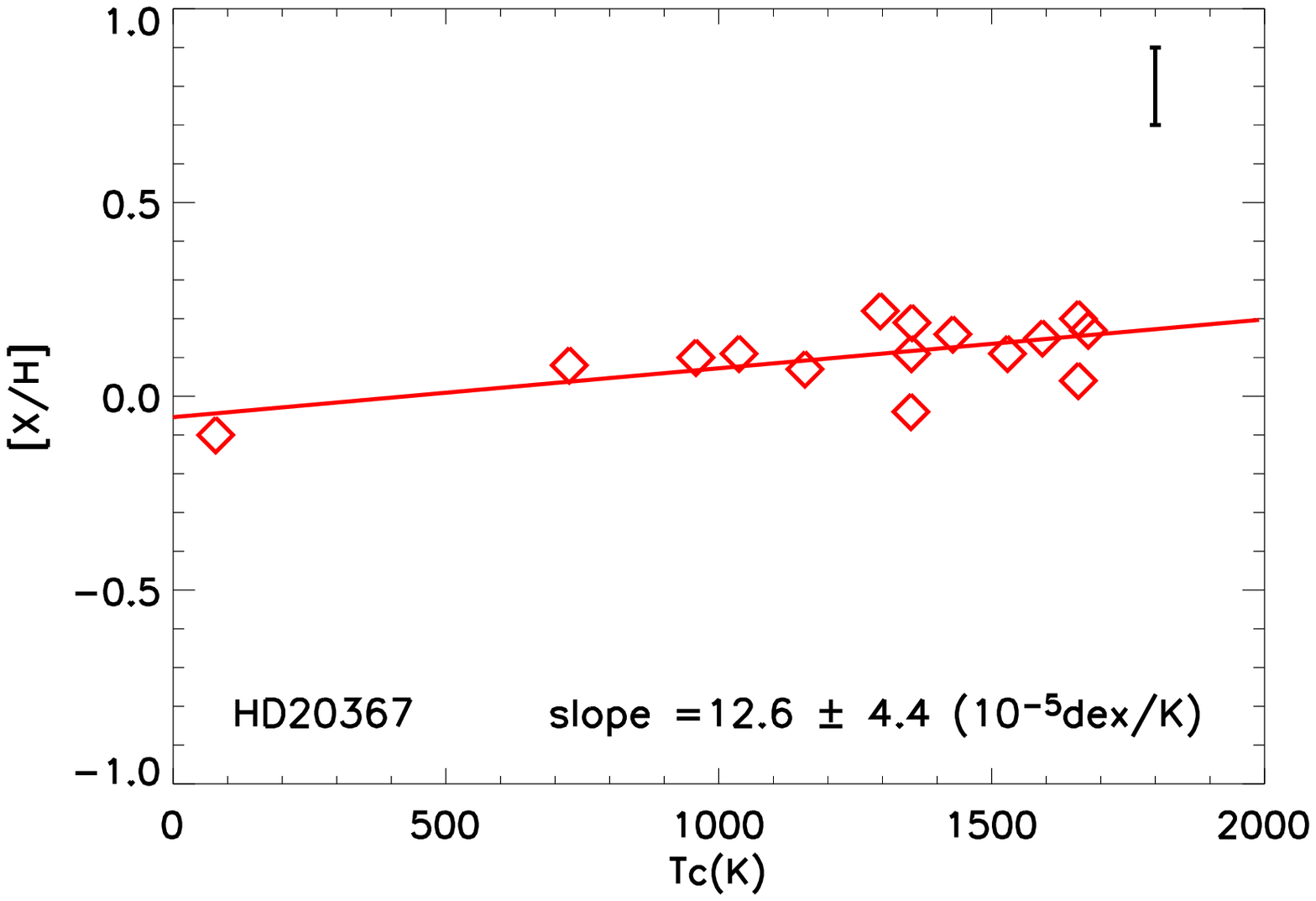}
\includegraphics[width=6.7cm]{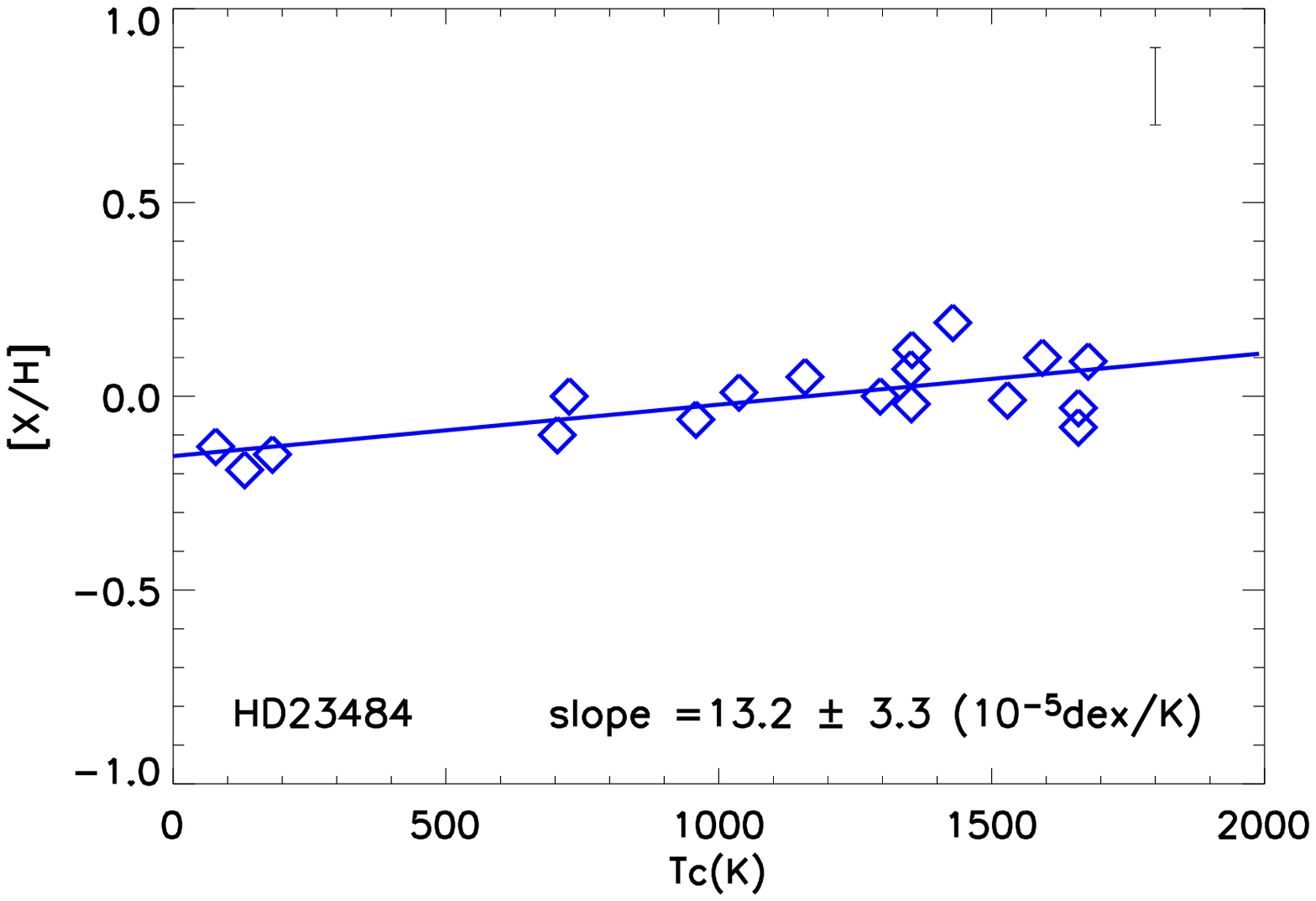}
\includegraphics[width=6.7cm]{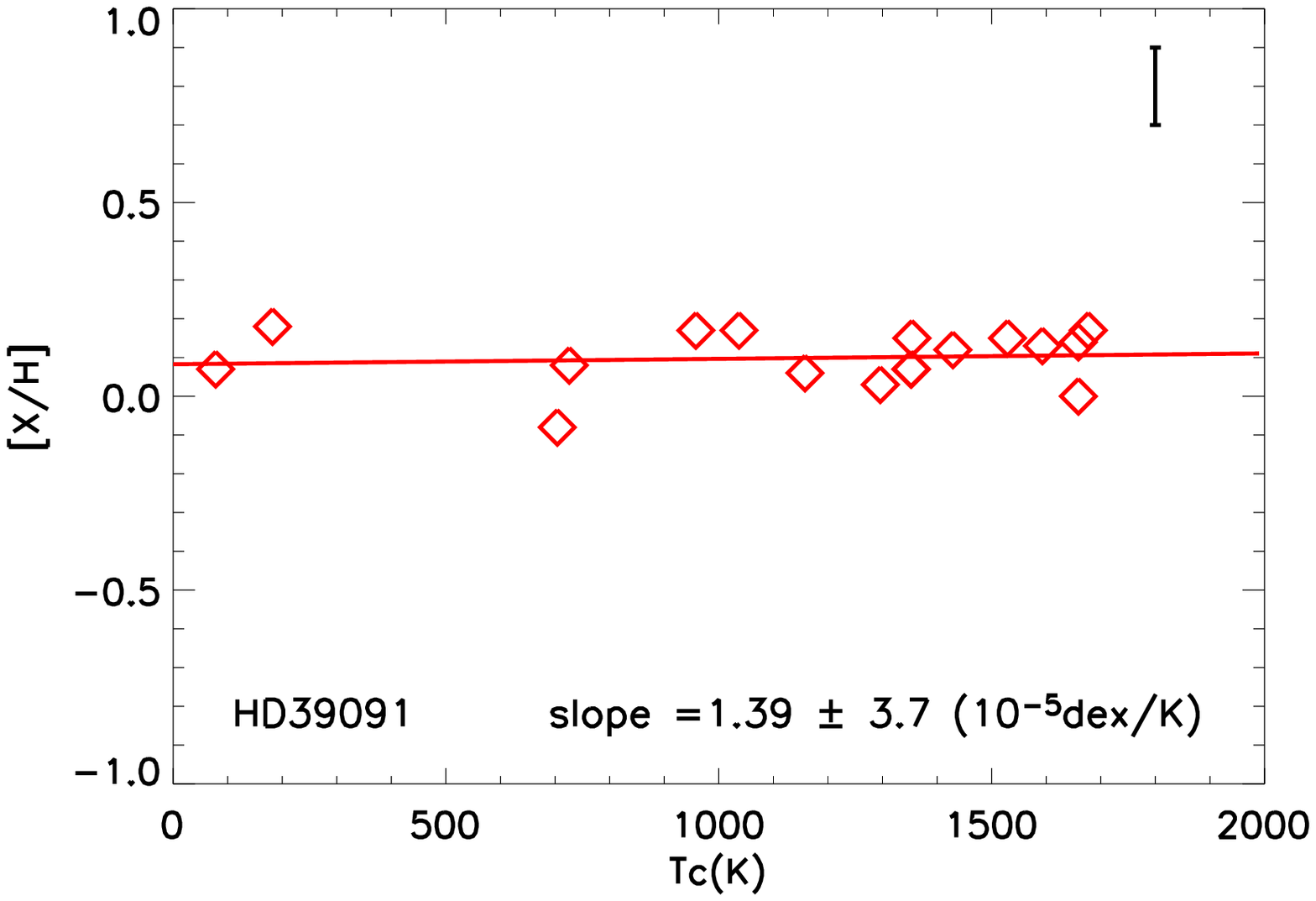}
\includegraphics[width=6.7cm]{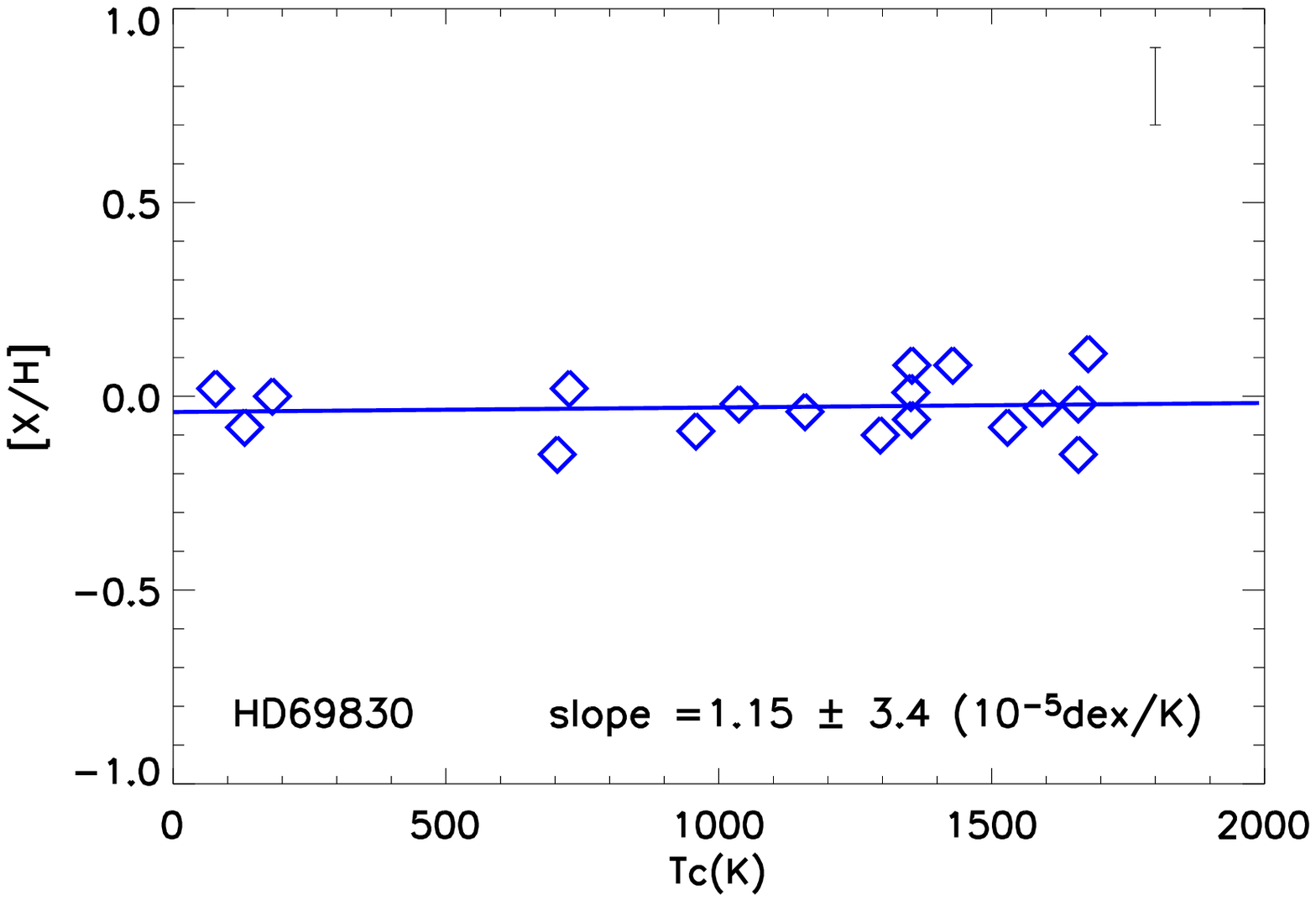}
\caption{Abundance ratios [$X$/H] plotted vs.\ the elemental condensation temperatures T$_C$ for the planet host \object{HD\,20367} and \object{HD\,39091} ({\it left panels}), and for the comparison sample stars \object{HD\,23484} and \object{HD\,69830} ({\it right panels}). The solid lines represent the linear least-squares fits to the data. The slopes values and the typical error bars associated to the [$X$/H] ratios are indicated at the bottom and the top of each plot, respectively.}
\label{Slope}
\end{figure*}

\begin{figure*}
\centering 
\includegraphics[width=12cm]{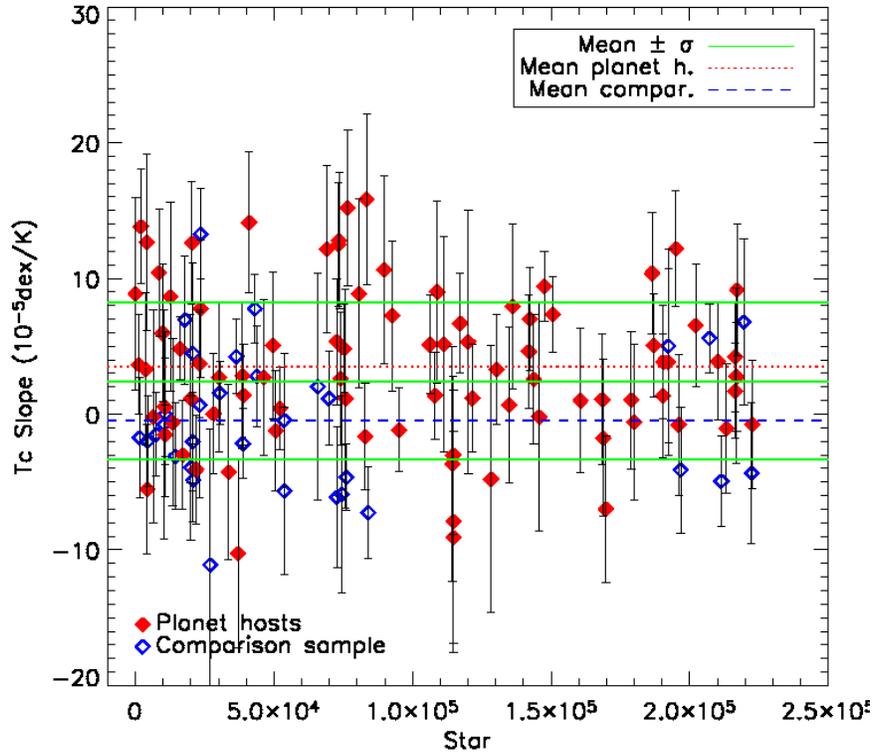}
\caption{Slopes derived from [$X$/H] vs. T$_C$ for all the planet host ({\it filled symbols}) and comparison sample ({\it open symbols}) stars. The solid lines represent the average slope of the two samples, and the average slope $\pm$ one standard deviation. The dotted and dashed lines indicate the average slope of stars with and without planets, respectively.}
\label{AllSlope}
\end{figure*}

\section{Introduction}
The announcement of the first planet orbiting around the solar type star \object{51Peg} by Mayor \& Queloz (\cite{May95}) marked the beginning of a steadily growing series of extrasolar planet discoveries. Radial velocity programmes have now found more than 150 planetary systems in the solar neighborhood. The study of all these planets and their stellar parents opens new opportunities to understand the mechanisms involved in planetary formation and evolution. Current analyses are giving the first statistically significant results about the properties of the new systems (e.g.\ Jorissen et al.\ \cite{Jor01}; Zucker \& Mazeh \cite{Zuc02}; Udry et al.\ \cite{Udr03}; Eggenberger et al.\ \cite{Egg04}).

The first strong link established between the planetary companions and their host stars is the fact that planet-harbouring stars are on average more metal-rich with respect to field stars. This idea has been put forward by Gonzalez (\cite{Gon97}), while the first clear evidence has been published by Santos, Israelian \& Mayor (\cite{San01}). Further studies have been confirming this result as new planet host candidates have been discovered (e.g. Gonzalez et al.\ \cite{Gon01}; Laws et al.\ \cite{Law03}; Santos et al.\ \cite{San03}, \cite{San04b}, \cite{San05}; Fischer \& Valenti \cite{Fis05}; for a review see Santos et al.\ \cite{San04a}).
This characteristic led to the suggestion that gas giant planet formation is favored by high stellar metallicity (Santos, Israelian \& Mayor \cite{San00}, \cite{San01}), so that planetary systems would be more likely to form out of metal-enriched primordial clouds. Alternatively, the metallic excess in these stars may be attributed to the pollution by the late ingestion of planetary material (Laughlin \cite{Lau00}, Gonzalez et al.\ \cite{Gon01}). 

Several results supporting the ``self-pollution" scenario were published. Murray \& Chaboyer (\cite{Mur02}) concluded that stochastic pollution by $\sim$5$M_{\oplus}$ of iron-rich material, together with selection effects and high intrinsic metallicity, may explain the observed metallicities in planet-harbouring stars. Israelian et al.\ (\cite{Isr01}, \cite{Isr03}) found evidence for a planet (planets or planetary material) having been engulfed by the discovery of a significant amount of $^6$Li in the stellar atmosphere of the parent star \object{HD\,82943}. However, the amount of accreted matter was not enough to explain the global stellar metallicity. Ingestion of planetary material may also explain the lithium and iron enhancement found by Gonzalez (\cite{Gon98}) and Laws \& Gonzalez (\cite{Law01}) in the primary component of the binary system 16 Cyg.
However, most studies today suggest that a primordial origin is much likelier to explain the metallicity excess in planet host stars. Pinsonneault et al.\ (\cite{Pin01}) ruled out the ``self-pollution'' hypothesis since the iron excess did not show the expected T$_{\rm eff}$ dependence. However, some different premises proposed by Vauclair (\cite{Vau04}) might invalidate their argument. An additional point in favour of the primodial scenario is the fact that the frequency of planets is a rising function of [Fe/H] (Santos et al.\ \cite{San01}, \cite{San04b}; Reid \cite{Rei02}). Moreover, despite of their huge convective envelopes, giants with planets do not present [Fe/H] lower than other planet hosts. We cannot thus completely rule out any of the different hypotheses proposed to explain the metallicity excess in planet host stars. Note also the strange behaviour reported for [$\alpha$/Fe] at supersolar [Fe/H], hard to be explained by Galactic chemical evolution (e.g. Bodaghee et al.\ \cite{Bod03}; Gilli et al.\ \cite{Gil05}; for a review see Israelian \cite{Isr04a}, \cite{Isr05}).

The abundance analyses of elements other than iron may give clues to this open question. Light elements are very important tracers of the internal structure and history of solar-type stars and therefore they can help to distinguish between different planet formation theories (Sandquist et al.\ \cite{Sand02}; Santos et al.\ \cite{San02}, \cite{San04c}; Israelian et al.\ \cite{Isr04b}). Volatile (with lower condensation temperatures T$_C$) and refractory (with higher condensation temperatures T$_C$) elements can also give information about the role played by pollution events on the global stellar chemical composition. In fact, the elements of the former group are expected to be deficient in accreted materials relative to the latter. If the infall of large amounts of rocky planetary material was the main cause of the metallicity excess in planet host stars, as the ``self-pollution'' scenario claims, an overabundance of refractory elements with respect to volatiles should be observed. This would imply an increasing trend of abundance ratios [X/H] with the elemental condensation temperature T$_C$. However, the engulfment of a whole planet (or the rapid infall of planetary material) may avoid the evaporation of volatile elements before being inside the star, leading to no peculiar trends of the stellar abundances with T$_C$.        

Smith, Cunha, \& Lazzaro (\cite{Smi01}) reported that a small subset of stars with planets exhibited an increasing [X/H] trend with T$_C$ and concluded that this trend pointed to the accretion of chemically fractionated solid material into the outer convective layers of these solar-type stars. They made use of the abundance results of 30 stars with planets reported by Gonzalez et al.\ (\cite{Gon01}) and Santos et al.\ (\cite{San01}), and compared them to those of 102 field stars from Edvardsson et al.\ (\cite{Edv93}) and Feltzing \& Gustafsson (\cite{Fel98}).  
Takeda et al.\ (\cite{Tak01})  also searched for a correlation between chemical abundances and T$_C$. They found that all volatile and refractories elements behave quite similarly in an homogeneously analysed set of 14 planet-harbouring stars and 4 field stars. Sadakane et al.\ (\cite{Sad02}) confirmed this result in 12 planet host stars, supporting a likelier primordial origin for the metal enhancement. Unfortunately, more results for planet-harbouring stars and an homogeneous comparison with field stars were needed to perform a more convincing test.

In this paper, we study the T$_C$ dependence of abundance ratios [$X$/H] uniformly derived in a large set of 105 planet host stars and in a volume-limited comparison sample of 88 stars without known planets. Some preliminary results were reported by Ecuvillon et al.\ (\cite{Ecu05b}). The large range of different T$_C$ covered by the analysed elements, which spans from 75 to 1600\,K, permits us to investigate possible anomalies in targets with planets beside comparison sample stars, and to detect hints of pollution. Our results offer new clues to understand the relative contribution of fractionated accretion to the metallicity excess observed in planet host stars.

\section{Data and Analysis}
We derived abundance ratios in an almost complete set of 105 planet host targets and in 88 comparison sample stars for the volatile elements CNO, S and Zn (Ecuvillon et al.\ \cite{Ecu04a}, \cite{Ecu04b},\cite{Ecu05a}) and the refractories Cu, Si, Ca, Sc, Ti, V, Cr, Mn, Co, Ni, Na, Mg and Al (Ecuvillon et al.\ \cite{Ecu04b}; Beirao et al.\ \cite{Bei05}; Gilli et al.\ \cite{Gil05}). All these abundances were computed using the homogeneous set of atmospheric parameters spectroscopically derived by Santos et al.\ (\cite{San04b}, \cite{San05}). We refer the reader to these papers for a description of the data and for the details of the spectral analysis. The uniform analyses applied to all the targets avoid possible errors due to differences in the line lists, atmospheric parameters, applied procedures, etc...  

For each target we obtained the [$X$/H] trend as function of the elemental condensation temperature T$_C$. The T$_C$ values for all the elements were taken from Lodders (\cite{Lod03}). We characterized each trend by computing the slope value corresponding to a linear least-squares fit, as proposed by Smith et al.\ (\cite{Smi01}). The targets with less than 14 abundance determinations were excluded from our study, in order to rely on the abundance ratios of at least two volatile elements. The selected targets, 88 planet host and 33 comparison sample stars, have [$X$/H] vs.\ T$_C$ trends spanning a large range of elemental condensation temperatures, from few dozen up to several hundreds kelvin, and thus leading to better constrained and reliable T$_C$ slopes. Figure~\ref{Slope} shows the [$X$/H] vs.\ T$_C$ trends for the targets \object{HD\,20367} and \object{HD\,39091}, and \object{HD\,23484} and \object{HD\,69830}, with and without planets, respectively. 

\begin{table}[!]
\caption[]{Targets with T$_C$ slopes larger than the average (2.4$\pm$5.8 10$^{-5}$\,dex\,K$^{-1}$) plus one standard deviation.}
\label{plus} 
\begin{center}
\begin{tabular}{lccr}
\hline
\noalign{\smallskip}
Object & Type & T$_C$ slope & [Fe/H] \\
 &  & (10$^{-5}$\,dex\,K$^{-1}$) & \\
\hline
\noalign{\smallskip}
\object{HD\,8574}  & plan &  10.43 &	0.06\\
\object{HD\,12661} & plan &   8.66 &	0.36\\
\object{HD\,40979} & plan &  14.13 &	0.21\\
\object{HD\,68988} & plan &  12.17 &	0.36\\
\object{HD\,73256} & plan &  12.48 &	0.26\\
\object{HD\,80606} & plan &   8.87 &	0.32\\
\object{HD\,83443} & plan &  15.84 &	0.39\\ 
\object{HD\,89744} & plan &  10.64 &	0.22\\ 
\object{HD\,108874}& plan &   9.00 &	0.23\\
\object{HD\,147513}& plan &   9.42 &    0.08\\
\object{HD\,186427}& plan &  10.38 &	0.08\\
\object{HD\,195019}& plan &  12.19 &	0.09\\
\object{HD\,217107}& plan &   9.14 &	0.37\\
\object{HD\,142}   & plan &   8.86 &	0.14\\
\object{HD\,2039}  & plan &  13.81 &	0.32\\
\object{HD\,4203}  & plan &  12.65 &	0.40\\
\object{HD\,20367} & plan &  12.62 &	0.17\\
\object{HD\,73526} & plan &  12.79 &	0.27\\
\object{HD\,76700} & plan &  15.19 &	0.41\\
\object{HD\,23484} & comp &  13.27 &	0.06\\
\noalign{\smallskip }		       
\hline  	    
\end{tabular} 	    
\end{center}	     
\end{table}

\begin{table}[!]
\caption[]{Targets with T$_C$ slopes lower than the average (2.4$\pm$5.8 10$^{-5}$\,dex\,K$^{-1}$) substracted one standard deviation.}
\label{minus} 
\begin{center}
\begin{tabular}{lccr}
\hline
\noalign{\smallskip}
Object & Type & T$_C$ slope & [Fe/H] \\
 &  & (10$^{-5}$\,dex\,K$^{-1}$) & \\
\hline
\noalign{\smallskip}
\object{HD\,22049}  & plan &$-$4.07 & $-$0.13\\
\object{HD\,37124}  & plan &$-$10.25& $-$0.38\\
\object{HD\,114762} & plan &$-$7.92& $-$0.70\\
\object{HD\,4208}   & plan &$-$5.52& $-$0.24\\
\object{HD\,33636}  & plan &$-$4.28& $-$0.08\\
\object{HD\,169830} & plan &$-$7.00&    0.21\\
\object{HD\,114386} & plan &$-$3.67&    0.04\\
\object{HD\,114783} & plan &$-$9.06&    0.09\\
\object{HD\,128311} & plan &$-$4.78&    0.03\\
\object{HD\,20010}  & comp &$-$3.94 & $-$0.19\\
\object{HD\,20807}  & comp &$-$4.86 & $-$0.23\\
\object{HD\,72673}  & comp &$-$6.12 & $-$0.37\\
\object{HD\,74576}  & comp &$-$5.90 & $-$0.03\\
\object{HD\,76151}  & comp &$-$4.66 &    0.14\\
\object{HD\,84117}  & comp &$-$7.25 & $-$0.03\\
\object{HD\,211415} & comp &$-$4.94 & $-$0.17\\
\object{HD\,222335} & comp &$-$4.35 & $-$0.16\\
\object{HD\,26965}  & comp &$-$11.10 & $-$0.31\\
\object{HD\,53706}  & comp &$-$5.67 & $-$0.26\\
\object{HD\,196761} & comp &$-$4.13 & $-$0.29\\
\noalign{\smallskip }		        
\hline  	    
\end{tabular} 	    
\end{center}	     
\end{table}

\begin{figure*}
\centering 
\includegraphics[width=12cm]{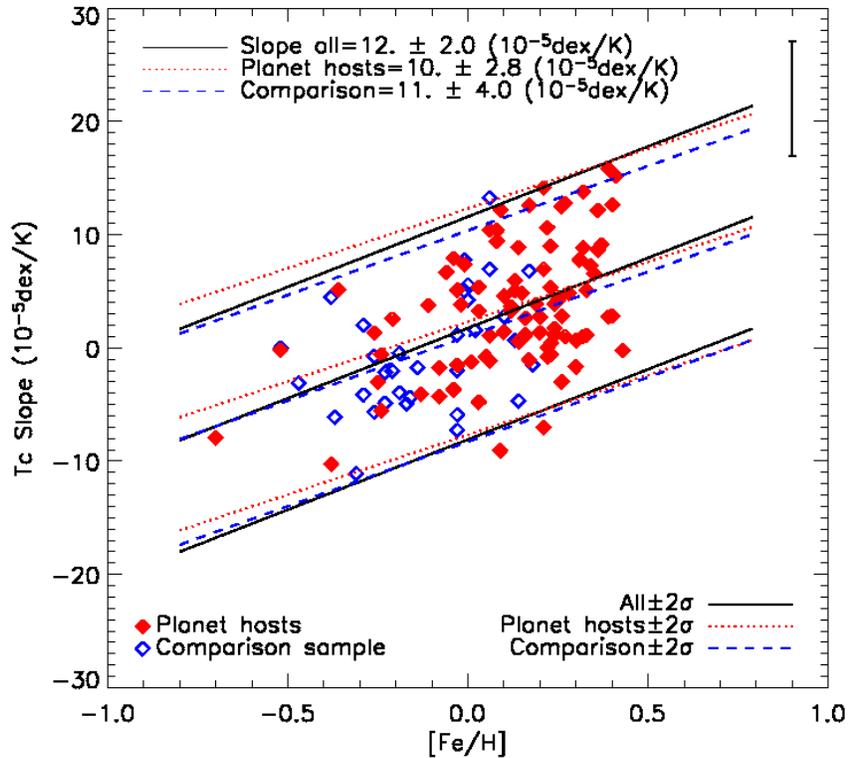}
\caption{Slopes from [$X$/H] vs. T$_C$ plotted vs.\ [Fe/H ]for all the planet host ({\it filled symbols}) and comparison sample ({\it open symbols}) stars. The solid lines represent the linear least-squares fit $\pm$\,2$\sigma$ for all the targets, while the dotted and dashed lines indicate the linear least-squares fits $\pm$\,2$\sigma$ for stars with and without planets, respectively. The slopes of each linear fit are shown at the bottom.}
\label{FeSlope}
\end{figure*}

\begin{figure*}
\centering 
\includegraphics[width=6.7cm]{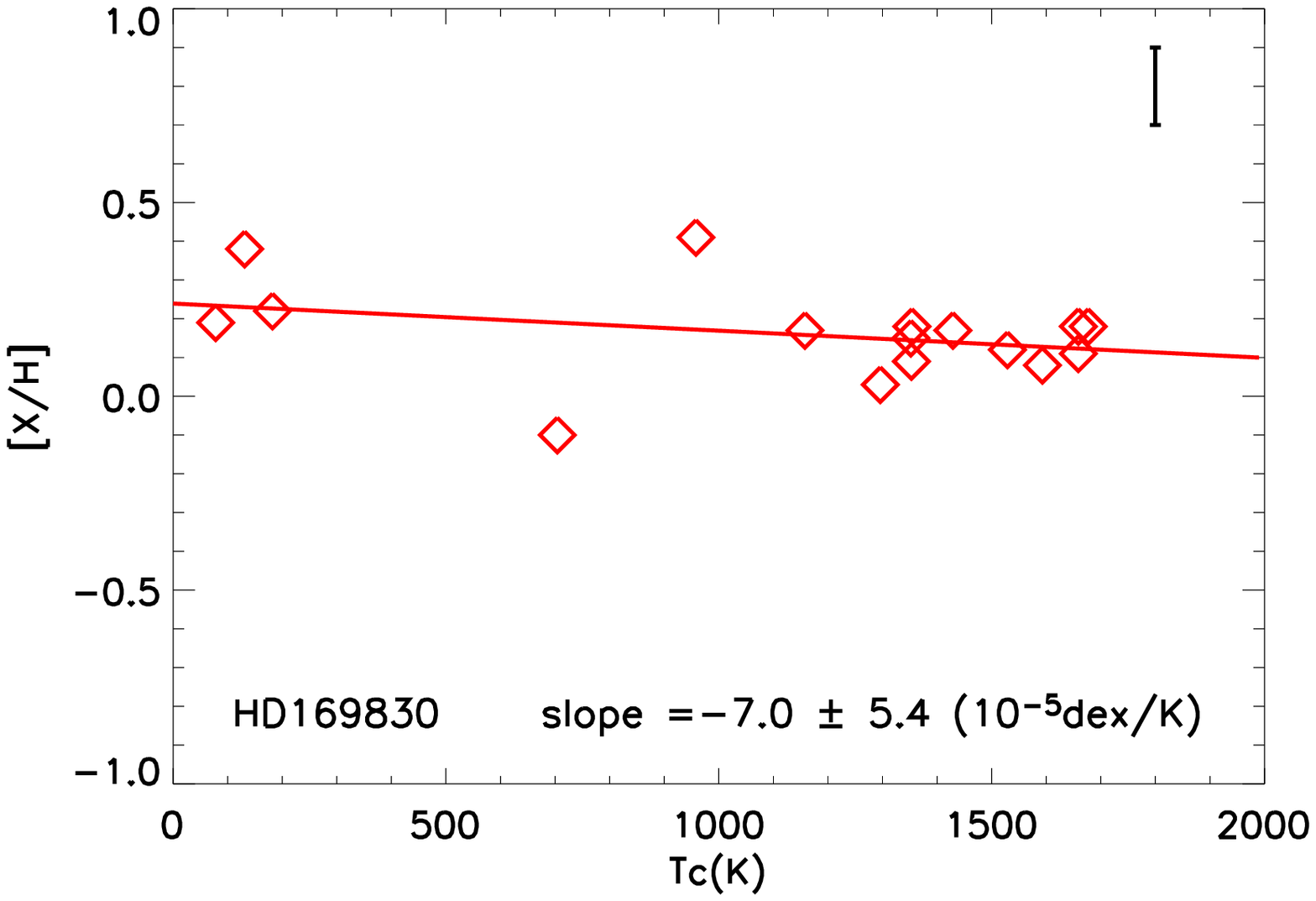}
\includegraphics[width=6.7cm]{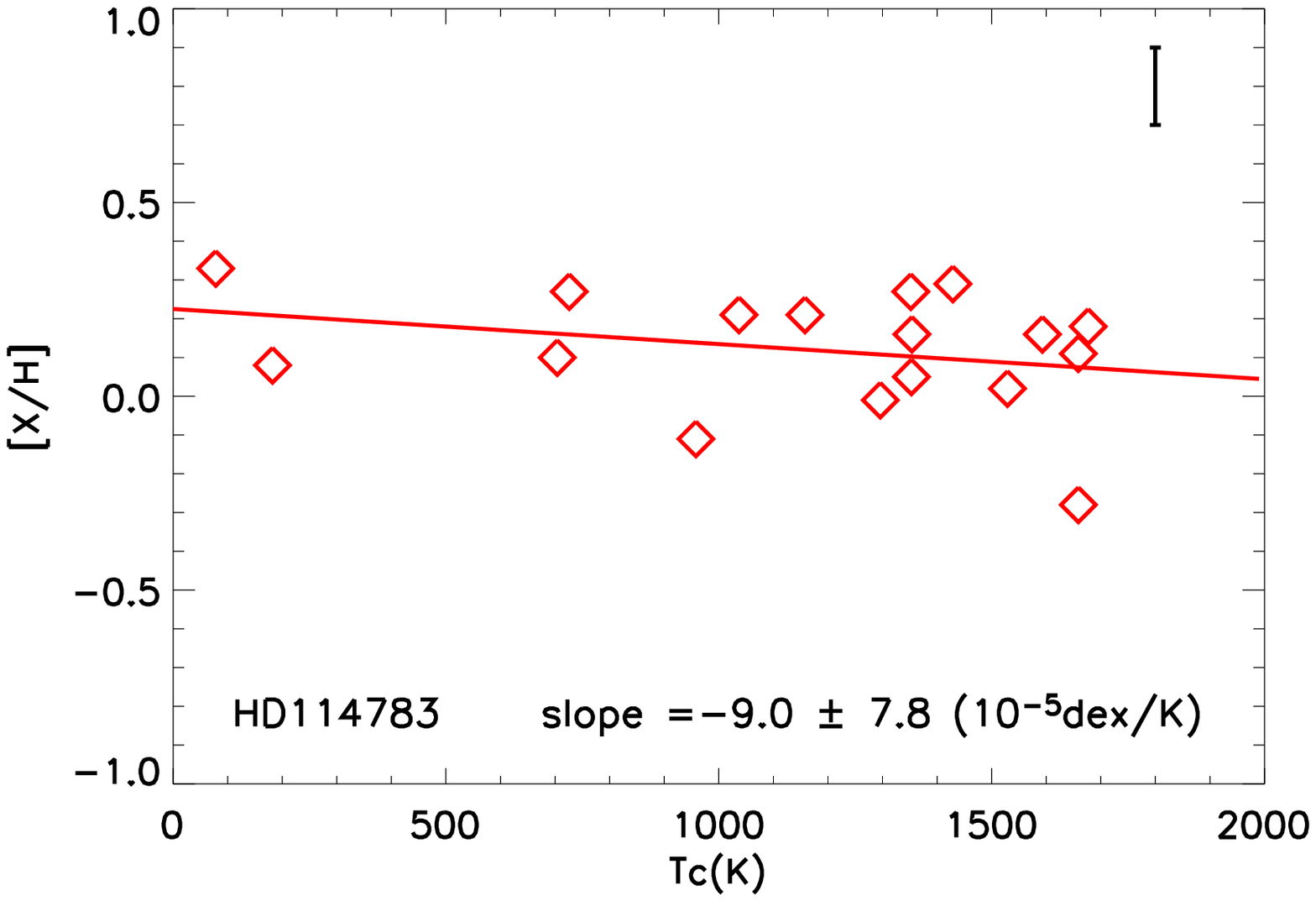}
\caption{Abundance ratios [$X$/H] plotted vs.\ the elemental condensation temperatures T$_C$ for the planet host stars \object{HD\,169830} and \object{HD\,114783}, having T$_C$ slopes more than 2\,$\sigma$ below the T$_C$ slopes vs.\ [Fe/H] fit. The solid lines represent the linear least-squares fits to the data. The slopes values and the typical error bars associated to the [$X$/H] ratios are indicated at the bottom and the top of each plot, respectively.}
\label{Targets}
\end{figure*}

\begin{table*}[!]
\caption[]{Targets with T$_C$ slopes more than 2\,$\sigma$ above the comparison sample fit. The atmospheric parameters are taken from Santos et al.\ (\cite{San04b}, \cite{San05}), while the metallic abundances are from Ecuvillon et al.\ (\cite{Ecu04a}, \cite{Ecu04b}, \cite{Ecu05a}) and Gilli et al.\ (\cite{Gil05}). The Li and Be abundances are those from Israelian et al.\ (\cite{Isr04b}) and Santos et al.\ (\cite{San04c}).}
\label{2sigmaplus} 
\begin{center}
\begin{tabular}{lccccccc}
\hline
\noalign{\smallskip}
 & \object{HD\,20367} & \object{HD\,23484} & \object{HD\,40979} & \object{HD\,76700} & \object{HD\,83443} & \object{HD\,195019} \\
\hline
\noalign{\smallskip}
Type & plan & comp & plan & plan & plan & plan \\
T$_C$ slope (10$^{-5}$\,K$^{-1}$) & 12.62 & 13.27  & 14.13 & 15.19 & 15.84 & 12.19	\\
T$_{\rm eff}$ (K)& 6138$\pm$79 & 5176$\pm$45 & 6145$\pm$42 & 5737$\pm$34 & 5454$\pm$61 & 5842$\pm$35 \\
$\log{g}$  & 4.53$\pm$0.22 & 4.41$\pm$0.17 & 4.31$\pm$0.15 & 4.25$\pm$0.14 & 4.33$\pm$0.17 & 4.32$\pm$0.07 \\
$\rm [Fe/H]$ & 0.17$\pm$0.10 & 0.06$\pm$0.05 & 0.21$\pm$0.05 & 0.41$\pm$0.05 & 0.35$\pm$0.08 & 0.08$\pm$0.04 \\
$\rm [C/H]$ & $-$0.10$\pm$0.09 & $-$0.13$\pm$0.10 &	0.10$\pm$0.07 &	0.21$\pm$0.07 & 0.35$\pm$0.08 & 0.06$\pm$0.07 \\
$\rm [N/H]$ & - & $-$0.19$\pm$0.12 & 0.10$\pm$0.16 & 0.32$\pm$0.14 & 0.26$\pm$0.14 & - \\
$\rm [O/H]$ & - & $-$0.15$\pm$0.11 &$-$0.18$\pm$0.08 & - & 0.15$\pm$0.09 &$-$0.20$\pm$0.14	\\
$\rm [S/H]$ & - &$-$0.10$\pm$0.09 & 0.00$\pm$0.08 &	0.10$\pm$0.08 &  0.52$\pm$0.10 &$-$0.15$\pm$0.06 \\ 
$\rm [Zn/H]$ & 0.08$\pm$0.11 & 0.00$\pm$0.08 &$-$0.04$\pm$0.12 & - & - & 0.02$\pm$0.06 \\ 
$\rm [Cu/H]$ & 0.11$\pm$0.11 & 0.01$\pm$0.09 & 0.01$\pm$0.09 & - & - & 0.12$\pm$0.07 \\
$\rm [Si/H]$ & 0.11$\pm$0.08 & $-$0.01$\pm$0.04 & 0.22$\pm$0.05 & 0.38$\pm$0.03 & 0.44$\pm$0.06 & 0.09$\pm$0.05 \\
$\rm [Ca/H]$ & 0.04$\pm$0.12 & $-$0.08$\pm$0.11 & 0.15$\pm$0.07 & 0.29$\pm$0.10 & 0.19 $\pm$0.12 & $-$0.03$\pm$0.10	\\
$\rm [Sc/H]$ & 0.20$\pm$0.13 & $-$0.03$\pm$0.12 & 0.18$\pm$0.10 & 0.47$\pm$0.08 & 0.55$\pm$0.11 & 0.15$\pm$0.09 \\
$\rm [Ti/H]$ & 0.15$\pm$0.12 & 0.10$\pm$0.08 & 0.18$\pm$0.06 & 0.45$\pm$0.05 & 0.46$\pm$0.10 & 0.13$\pm$0.06 \\
$\rm [V/H]$  & 0.16$\pm$0.13 & 0.19$\pm$0.09 &0.24$\pm$0.13	& 0.53$\pm$0.08	& 0.61$\pm$0.12 & 0.12$\pm$0.06	\\
$\rm [Cr/H]$ & 0.22$\pm$0.09 & 0.00$\pm$0.05 & 0.18$\pm$0.10 & 0.36$\pm$0.08 & 0.33$\pm$0.08 & 0.03$\pm$0.06 \\
$\rm [Mn/H]$ & 0.07$\pm$0.14 & 0.05$\pm$0.09 & 0.17$\pm$0.16 & 0.51$\pm$0.12 &  0.58$\pm$0.16 & 0.12$\pm$0.05 \\
$\rm [Co/H]$ &$-$0.04$\pm$0.15 & 0.07$\pm$0.07 & 0.10$\pm$0.11 & 0.57$\pm$0.10 & 0.63$\pm$0.09 & 0.06$\pm$0.06 \\
$\rm [Ni/H]$ & 0.11$\pm$0.09 & $-$0.02$\pm$ 0.05 & 0.16$\pm$0.08 & 0.41$\pm$0.06 & 0.49$\pm$0.07 & 0.03$\pm$0.04 \\
$\rm [Na/H]$ & 0.10$\pm$0.05 & $-$0.06$\pm$0.08 & 0.34$\pm$0.04 & 0.40$\pm$0.06 &  - & 0.04$\pm$0.03 \\
$\rm [Mg/H]$ & 0.19$\pm$0.08 & 0.12$\pm$0.05 & 0.33$\pm$0.05 & 0.54$\pm$0.05 & 0.49$\pm$0.05 & 0.13$\pm$0.03 \\
$\rm [Al/H]$ & 0.17$\pm$0.05 & 0.09$\pm$0.04 & - & 0.52$\pm$0.04 & 0.53$\pm$0.06 & 0.19$\pm$0.02 \\ 
$\log{\epsilon}$(Li) & 3.02 & $<$0.44 & - & - & $<$0.56 & 1.46 \\
$\log{\epsilon}$(Be) & - & $<$0.70 & - & - & $<$0.70 & 1.15 \\  
\noalign{\smallskip }		       
\hline  	    
\end{tabular} 	    
\end{center}	     
\end{table*}

\begin{table}[!]
\caption[]{Targets with T$_C$ slopes more than 2\,$\sigma$ below the fit. The atmospheric parameters are taken from Santos et al.\ (\cite{San04b}, \cite{San05}), while the metallic abundances are from Ecuvillon et al.\ (\cite{Ecu04a}, \cite{Ecu04b}, \cite{Ecu05a}) and Gilli et al.\ (\cite{Gil05}). The Li and Be abundances are those published by Israelian et al.\ (\cite{Isr04b}) and Santos et al.\ (\cite{San04c}).}
\label{twosigmaminus}
\begin{center}
\begin{tabular}{lccc}
\hline
\noalign{\smallskip}
 & \object{HD\,114783} & \object{HD\,169830} \\
\hline
\noalign{\smallskip}
Type & plan & plan \\
T$_C$ slope (10$^{-5}$\,K$^{-1}$) & $-$9.07 & $-$7.00 \\
T$_{\rm eff}$ (K)& 5098$\pm$36 & 6299$\pm$41 \\
$\log{g}$ & 4.45$\pm$0.11 & 4.10$\pm$0.02 \\
$\rm [Fe/H]$ & 0.09$\pm$0.04 & 0.21$\pm$0.05 \\
$\rm [C/H]$ & 0.33$\pm$0.13 & 0.19$\pm$0.04 \\
$\rm [N/H]$ & - & 0.38$\pm$0.12 \\
$\rm [O/H]$ & 0.08$\pm$0.18 & 0.22$\pm$0.12 \\
$\rm [S/H]$ & 0.10$\pm$0.07 & $-$0.10$\pm$0.06 \\ 
$\rm [Zn/H]$ & 0.27$\pm$0.07 & - \\ 
$\rm [Cu/H]$ & 0.21$\pm$0.07 & - \\
$\rm [Si/H]$ & 0.02$\pm$0.06 & 0.12$\pm$0.05 \\
$\rm [Ca/H]$ & $-$0.28$\pm$0.07 & 0.11$\pm$0.07 \\
$\rm [Sc/H]$ & 0.11$\pm$0.08 & 0.18$\pm$0.07\\
$\rm [Ti/H]$ & 0.16$\pm$0.08 & 0.08$\pm$0.04 \\
$\rm [V/H]$  & 0.29$\pm$0.10 & 0.17$\pm$0.13 \\
$\rm [Cr/H]$ & $-$0.01$\pm$0.04 & 0.03$\pm$0.03 \\
$\rm [Mn/H]$ & 0.21$\pm$0.06 & 0.17$\pm$0.10 \\
$\rm [Co/H]$ & 0.27$\pm$0.10 & 0.15$\pm$0.08 \\
$\rm [Ni/H]$ & 0.05$\pm$0.09 & 0.09$\pm$ 0.05 \\
$\rm [Na/H]$ &$-$0.11$\pm$0.04 & 0.41$\pm$0.02 \\
$\rm [Mg/H]$ & 0.16$\pm$0.04 & 0.18$\pm$0.06 \\
$\rm [Al/H]$ & 0.18$\pm$0.06 & 0.18$\pm$0.03 \\
$\log{\epsilon}$(Li) & $-$0.12 & 1.16 \\
$\log{\epsilon}$(Be) & - & $<$-0.40 \\
\noalign{\smallskip }		       
\hline  	    
\end{tabular} 	    
\end{center}	      
\end{table}

\section{Results and Discussion}

The slopes derived from [$X$/H] vs.\ T$_C$ for all the planet host and comparison sample stars are shown in Figure~\ref{AllSlope}. Error bars reflect the statistical uncertainty in the derived slopes. A subsample of 19 and 1 stars with and without planets, respectively, stands out for having slopes larger than the average plus one standard deviation (see Table~1). The slopes are such that species with low condensation temperatures are 0.2\,dex lower in abundance than the more refractory species. These targets might be candidate for having possibly been polluted by the ingestion of fractionated material (see the [$X$/H] vs.\ T$_C$ trends for \object{HD\,20367} and \object{HD\,23484} in Figure~\ref{Slope}). It is important to note that this subgroup shows in average a higher metallicity, of the order of 0.2\,dex.\\ 
Another group of 9 planet hosts and 11 comparison sample stars emerges from the same representation because of their slopes lower than the average substracted one standard deviation (see Table~2). These targets correspond to a quite low average metallicity of the order of $-$0.2\,dex. This characteristic, together with the similar one pointed out for the subgroup with higher slopes, suggests that this behaviour is related with the underlying chemical evolutionary effects. The fact that a large number of comparison sample stars behaves in the same way as planet host stars adds a further argument to the consideration that galactic chemical evolution, instead of fractionated accretion, sets the pace for this representation. 

\subsection{T$_C$ slopes as function of [Fe/H]}
\label{Fe}
In order to disentangle the effects due to the chemical evolution from those related to the fractionated accretion, we analysed the dependence of T$_C$ slopes on the stellar metallicity. Figure~\ref{FeSlope} shows the T$_C$ slopes vs.\ [Fe/H] and the linear least-squares fits corresponding to all the targets, and to the two samples of stars with and without planets, independently. All the linear least-squares fits were obtained by weighting each point for its statistical uncertainty. There is a clear global trend of slopes to increase with metallicity, which is an expected signature of the Galactic chemical evolution. In fact, [C/Fe] and [O/Fe] rise steeply toward lower metallicities (e.g.\ Gustafsson et al.\ \cite{Gus99}; Ecuvillon et al.\ \cite{Ecu04b}, \cite{Ecu05a}; Takeda \& Honda \cite{Tak05}), and they thus tend to produce negative slopes in [$X$/H] vs.\ T$_C$ for targets with [Fe/H]$<$0.1.\\
The set of planet host stars seems to behave quite similarly to the comparison sample. Both groups present similar dispersions and slopes, even if the number of included comparison sample stars (33) is much lower than planet host targets (88). However, the fit corresponding to the planet host set is shifted towards higher T$_C$ slope values. By visual inspection, there seems to appear a bulk of 10 planet host and a comparison sample stars with particularly high T$_C$ slopes at supersolar metallicities, which fall above the general scatter. Their abundance pattern is slightly enhanced with respect to that of the rest of stars probably due to the general chemical evolutionary effects. These targets may be the main responsible for the shift observed in the fit of planet host stars, and they could enclose possible signature of selective accretion.\\
In order to explore this possibility, we checked which targets fall out from the limits traced by the fits $\pm$2\,$\sigma$. 
Due to the much larger number of stars with planets than comparison sample stars, the mean trend of both samples together is strongly marked by the behaviour of the targets with planets, especially at high metallicities. The fit of the two samples together is hence shifted toward high T$_C$ slope values, in the same way as the fit of planet hosts is (see the solid and dotted lines in Figure~\ref{FeSlope}). In fact, only the target \object{HD\,23484} belonging to the comparison sample presents a T$_C$ slope more than 2\,$\sigma$ above the global fit. Other two targets with known planets, \object{HD\,114783} and \object{HD\,169830}, stand out for having T$_C$ slopes more than 2\,$\sigma$ below the global fit. All these targets were already reported in Tables~1 and~2 for having T$_C$ slopes more than one standard deviation away from the average. The atmospheric parameters and chemical abundances of these peculiar stars are listed in Table~3 and Table~4, while their [$X$/H] vs.\ [Fe/H] trends are shown in Figure~\ref{Slope} (\object{HD\,23484}, {\it top right panel}) and Figure~\ref{Targets} (\object{HD\,114783} and \object{HD\,169830}).\\
An interesting feature is that several planet host stars show slopes more than 2\,$\sigma$ above the fit corresponding to the comparison sample. In particular, T$_C$ slopes more than 2\,$\sigma$ above the fit of the comparison sample are observed in the comparison sample target \object{HD\,23484} and in 5 planet host stars, all previously reported in Table~1. All the atmospheric parameters and chemical abundances of these peculiar stars are listed in Table~3. The same two planet host stars previously  reported for having slopes more than 2\,$\sigma$ below the global fit, \object{HD\,114783} and \object{HD\,169830}, fall also more than 2\,$\sigma$ below the fit of the comparison sample (see Table~4).\\
It is important to stress that the global trend is heavily influenced by the behaviour of the planet host stars. If a subset of planet host stars bore some signature of fractionated accretion, the global trend would somehow enclose the same effect. This would affect the resulting fit to a greater extent if the pollution sign was more evident in metal-rich objects, since the lack of comparison sample targets drastically increases at supersolar metallicities. In this framework, the subsample of planet host stars standing more than 2\,$\sigma$ above the linear least-squares fit traced by the comparison sample might be interpreted as tentative candidates for having been strongly polluted by fractionated material and whose photospheres still keep the signature of these events. The fact that these same objects do not fall 2\,$\sigma$ out the global fit could be explained by the great importance that planet host stars at high metallicities exert on the fit. However, these suggestions are only tentative, and the possibility of accretion signatures in these targets must be confirmed by other pieces of information, such as detailed abundance studies of light elements, and in particular of the isotopic ratio $^6$Li/$^7$Li.\\
When available, the abundances of Li and Be abundances were extracted from the works by Israelian et al.\ (\cite{Isr04b}) and Santos et al.\ (\cite{San04c}) and listed in Tables~3 and~4. Although the planet host target \object{HD\,20367} stands out for having very high Li content, this value is completely ``normal'' in main-sequence stars with T$_{\rm eff}$ above 6000\,K, which preserve a significant fraction of their original lithium. The Li deficiency in \object{HD\,83443}, \object{HD\,23484} and \object{HD\,114783} is not surprising either, because of their T$_{\rm eff}$ lower than 5500\,K. The planet host \object{HD\,195019} presents the significant Li depletion reported by Israelian et al.\ (\cite{Isr04b}) in solar-type stars with planets. Some mechanism might exist which makes stars with planets more efficient in depleting lithium than ``single'' stars. However, this result does not contribute in any extent to confirm that our targets keep signature of fractionated accretion. Moreover, none of these targets was reported by these authors for presenting any anomalous correlation of the lithium content with the orbital parameters.

\begin{figure*}
\centering 
\includegraphics[width=6.7cm]{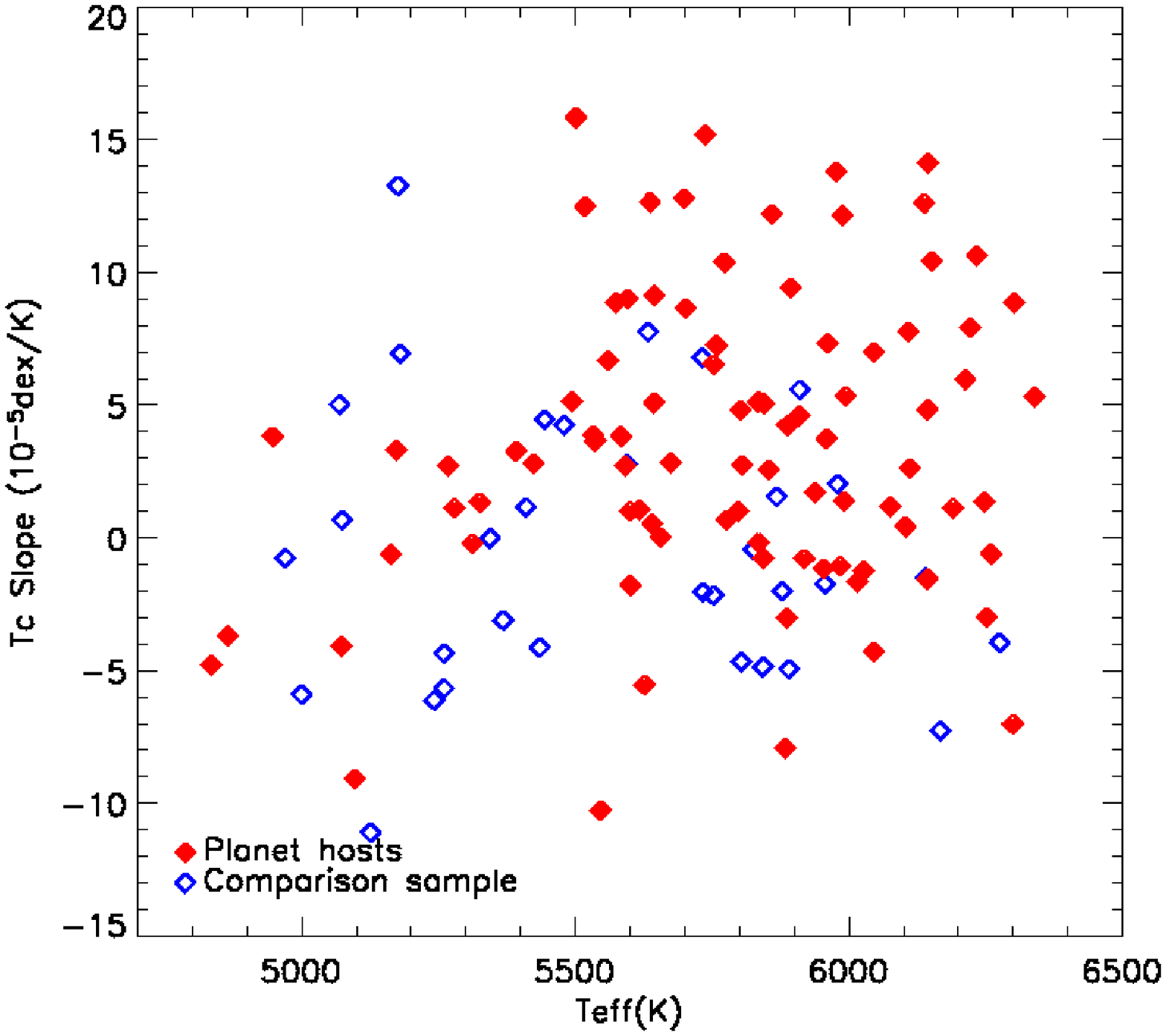}
\includegraphics[width=6.7cm]{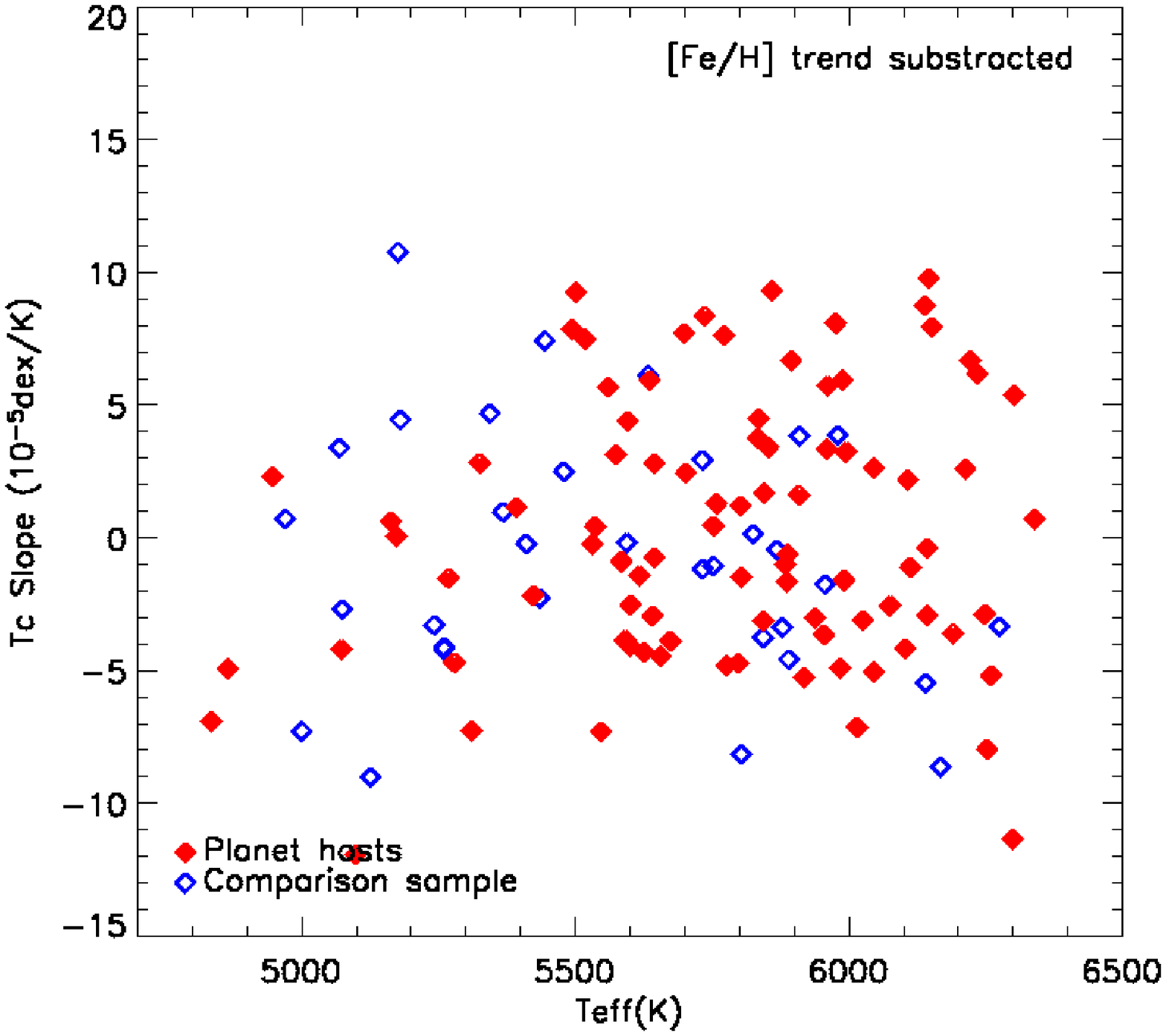}
\caption{Slopes from [$X$/H] vs. T$_C$ plotted vs.\ T$_{\rm eff}$ for all the planet host ({\it filled symbols}) and comparison sample ({\it open symbols}) stars.}
\label{TeffSlope}
\end{figure*}

\begin{figure*}
\centering 
\includegraphics[width=12cm]{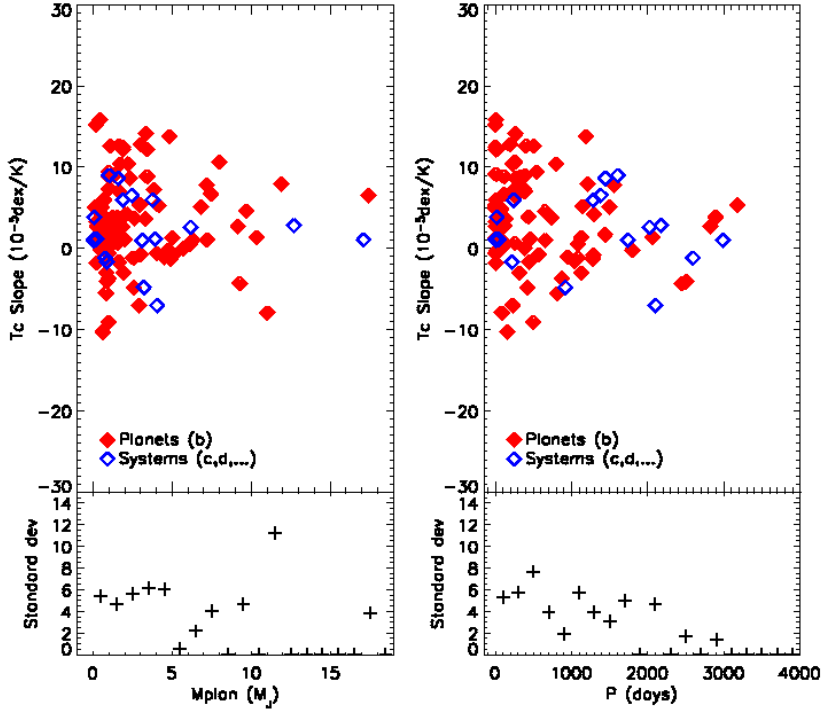}
\caption{Slopes from [$X$/H] vs. T$_C$ plotted vs.\ the planetary mass in M$_J$ units ({\it left panel}) and the planetary period in days ({\it right panel}) for all the planets: b components ({\it filled symbols}) and c,d,... components of planetary systems ({\it open symbols}). The lower panels show the standard deviations per bin corresponding to the above representations.}
\label{MjPSlope}
\end{figure*}

\begin{figure*}
\centering 
\includegraphics[width=12cm]{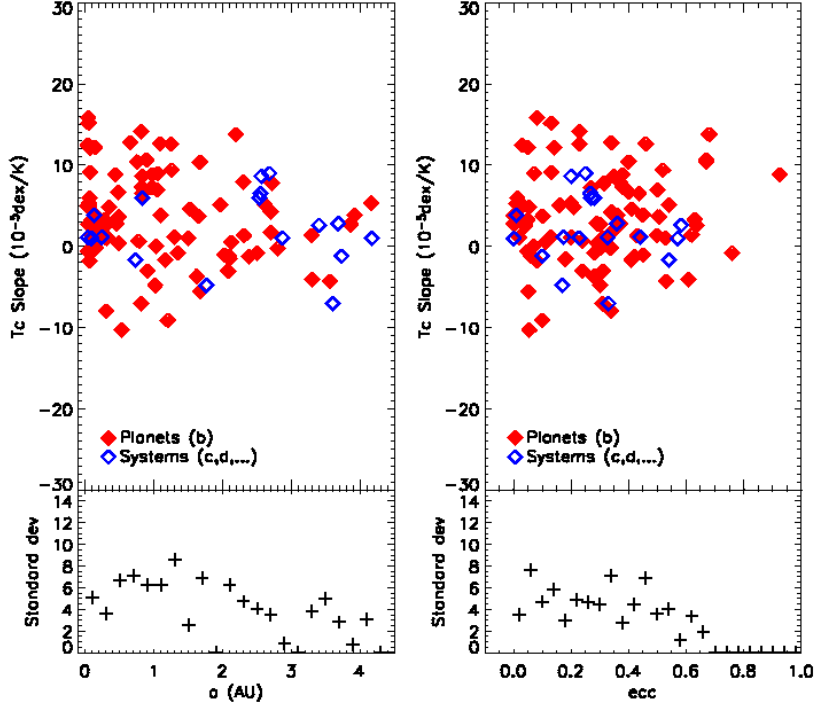}
\caption{Slopes from [$X$/H] vs. T$_C$ plotted vs.\ the orbital semi-major axis in AU ({\it left panel}) and the orbital eccentricity ({\it right panel}) for all the planets: b components ({\it filled symbols}) and c,d,... components of planetary systems ({\it open symbols}).  The lower panels show the standard deviations per bin corresponding to the above representations.}
\label{aEccSlope}
\end{figure*}

\begin{table}[!]
\caption[]{Planetary parameters of targets with T$_C$ slopes more than 2\,$\sigma$ above the comparison sample fit, from Extrasolar Planets Encyclopaedia (http://www.obspm.fr/encycl/encycl.html) compiled by Jean Schneider. }
\begin{center}
\begin{tabular}{lccccccc}
\hline
\noalign{\smallskip}
Planet & Mass (M$_J$) & Period (days) & a (AU) & Ecc. \\
\hline
\noalign{\smallskip}
\object{HD\,20367}b & 1.1 & 500 & 1.25 & 0.23 \\
\object{HD\,40979}b & 3.3 & 267 & 0.81 & 0.23 \\ 
\object{HD\,76700}b & 0.2 & 4 & 0.05 & 0.13 \\
\object{HD\,83443}b & 0.4 & 3 & 0.04 & 0.08 \\
\object{HD\,195019}b & 3.4 & 18 & 0.14 & 0.05 \\
 \noalign{\smallskip }		       
\hline  	    
\end{tabular} 	    
\end{center}	    
\label{planparabove}  
\end{table}

\begin{table}[!]
\caption[]{Planetary parameters of targets with T$_C$ slopes more than 2\,$\sigma$ below the comparison sample fit, from Extrasolar Planets Encyclopaedia (http://www.obspm.fr/encycl/encycl.html) compiled by Jean Schneider. }
\begin{center}
\begin{tabular}{lccccccc}
\hline
\noalign{\smallskip}
Planet & Mass (M$_J$) & Period (days) & a (AU) & Ecc. \\
\hline
\noalign{\smallskip}
\object{HD\,114783}b & 1.0 & 501 & 1.2 & 0.1 \\
\object{HD\,169830}b & 2.9 & 226 & 0.8 & 0.3 \\
\noalign{\smallskip}		       
\hline  	    
\end{tabular} 	    
\end{center}	    
\label{planparbelow}  
\end{table}
  
\subsection{T$_C$ slopes as function of T$_{\rm eff}$}
If pollution of the outer layers by infall is important, then also a trend with T$_{\rm eff}$ should be observed, since the convection zone decreases rapidly with increasing mass -- and T$_{\rm eff}$ -- for main-sequence stars. Pollution effects are expected to be much more evident in stars with T$_{\rm eff}>$ 6000\,K, where the surface convective zones have little mass and even a little amount of accretion would result in apparent metallic abundances. Figure~\ref{TeffSlope} ({\it left panel}) shows the slopes derived from [$X$/H] vs.\ T$_C$ as function of T$_{\rm eff}$. The two groups of stars with and without planets display considerable scatter, but it seems to appear a large number of planet host stars with high T$_C$ slope values for T$_{\rm eff}$ above 5500\,K. This would be consistent with the behaviour predicted by the ``self-pollution'' hypothesis. Nevertheless, we have to take into account that these slopes are affected by the trend with [Fe/H] due to Galactic chemical evolutionary effects. In order to clear the representation of T$_C$ slopes vs.\ T$_{\rm eff}$  from this component, we substracted the linear trend with [Fe/H] from the data (see Figure~\ref{TeffSlope}, {\it right panel}), as proposed by Gonzalez (\cite{Gon03}). The results do not show the previously observed incidence of planet host stars with high T$_C$ slope values at $T_{\rm eff}>$ 5500\,K. Stars with planets present T$_C$ slopes spanning the same range of values than the comparison sample with a similar scatter. The only noticeable difference is that the T$_{\rm eff}$ distribution of the set of stars with planets is shifted toward higher T$_{\rm eff}$. This is due to the fact that our comparison sample is particularly rich in cooler stars. This results suggests that the fractionated pollution may not be the main cause of the chemical anomalies found in planet host stars, and would support a ``primordial'' origin as much likelier. 

\subsection{T$_C$ slopes as function of planetary parameters}  
The physical parameters of exoplanets can give new hints to investigate the importance of pollution events on the chemical abundances of the planet host targets studied in this work. Some studies (Gonzalez \cite{Gon98}; Queloz et al.\ \cite{Que00}; Sozzetti \cite{Soz04}) reported a possible relation between the stellar metallicity and the orbital separation, observing that stars hosting hot Jupiters tended to be more metal-rich than the rest. However, Santos et al.\ (\cite{San01}, \cite{San03}) did not find such a significant trend. Smith et al.\ (\cite{Smi01}) found that the subgroup of stars reported for bearing possible accretion signatures standed out as having smaller orbital separations, and possibly smaller eccentricities and companion masses. A recent work by Sozzetti (\cite{Soz04}) investigated the correlation between the stellar metallicity and the orbital period. It reported the absence of very short-period planets around stars with [Fe/H]$<$0 as a possible evidence of the metallicity dependence of the migration rates of giant planets in protoplanetary discs.\\
In Figure~\ref{MjPSlope} and Figure~\ref{aEccSlope} the stellar [$X$/H] vs.\ T$_C$ slopes are represented as function of the planetary mass, orbital period, semi-major axis and eccentricity. The parameters of planets were obtained from the Extrasolar Planets Encyclopaedia (http://www.obspm.fr/encycl/encycl.html) compiled by Jean Schneider. No particular correlation emerges between the slope values and any planetary parameters. By visual inspection, the dispersion of T$_C$ slopes seems to decrease with the planetary mass and period (see Figure~\ref{MjPSlope}). However, we computed the standard deviation per bin to quantify the dispersion and visualize its behaviour as function of the parameter (see {\it lower panels}), and no peculiar trends appeared. The observed distributions are simply related to the number of planets with a given value of the planetary parameter.\\ 
All the planetary parameters corresponding to the peculiar stars with slopes more than 2\,$\sigma$ above and below the fits are listed in Tables~5 and~6, respectively. Comparing with the distributions of all the analysed planet host stars, these objects are orbited by low mass planetary companions at short periods. \object{HD\,76700}, \object{HD\,83443} and \object{HD\,195019} have planets orbiting with particularly short periods and separations, what makes the infall of large amounts of planetary material much likelier to occur. Moreover, smaller orbital separations suggest the possibility of more interaction between planet, disk, and star, as the planet presumably migrated inward to its current position around its parent star when it was forming. 
All this supports the suggestion that these stars bear the signature of fractionated pollution.      

\section{Conclusions}
After gathering detailed and uniform abundance ratios of volatile and refractory elements in a large set of planet host stars and a volume-limited comparison sample of stars without any known planets, we derived the [$X$/H] vs.\ T$_C$ trend and the slope value corresponding to the linear fit for each target. Planet host stars present an average slope higher than the comparison sample. However, this characteristic is mainly due to chemical evolutionary effects, more evident in planet host stars because of their metal-rich nature.
The obtained increasing trend of T$_C$ with metallicity is a consequence of the Galactic chemical evolution. There does not seem to appear any remarkable difference in the behaviour of stars with and without planetary companions. However, a subset of 5 planet host stars and 1 comparison sample target with slopes falling out the trends are proposed as possible candidates to exhibit the signature of fractionated accretion. The larger number of planet host stars with respect to the comparison sample and the statistical uncertainties affecting the slope values do not allow a conclusive interpretation. \\
Further evidence of pollution are investigated by looking for any possible dependence on T$_{\rm eff}$ and planetary parameters. No clear trends emerge with the stellar T$_{\rm eff}$, contrary to the results published by Gonzalez (\cite{Gon03}). We did not observe any significant behaviour with the planetary mass, orbital period, separation and eccentricity. Similar results were obtained by Santos et al.\ (\cite{San03}) when looking for possible correlations between the physical parameters of the exoplanets and the metallicity excess of their parent stars. Some of the targets we reported as candidates for possible selective accretion show planetary parameters compatible with a ``self-pollution'' scenario.\\
In conclusion, these possible candidates for self-pollution have to be carefully analyzed and submitted to further tests, in order to confirm or reject this suggestion. On the whole, our results do not point to a solely ``primordial'' or ``self-pollution'' scenario to explain the observed trends. Although in most cases a mainly primordial origin of the metallic excess in planet host stars seems likelier, probably a more complex mechanism combining both scenarios may underlie the observations.    

\begin{acknowledgements}
Support from Funda\c{c}\~ao para a Ci\^encia e a Tecnologia (Portugal) to N.C.S. in the form of a scholarship (reference  SFRH/BPD/8116/2002) and a grant (reference POCI/CTE-AST/56453/2004) is gratefully acknowledged. We thank the anonymous referee for useful comments.
\end{acknowledgements}


\begin{thebibliography}{} 

\bibitem[2005]{Bei05}
Beir\~ao, P., Santos, N. C., Israelian, G., Mayor, M. 2005, A\&A, 438, 251

\bibitem[2003]{Bod03}
Bodaghee, A., Santos, N. C., Israelian, G., Mayor, M. 2003, A\&A, 404, 715

\bibitem[2004a]{Ecu04a}
Ecuvillon, A., Israelian, G., Santos, N. C., Mayor, M., Garc\'{\i}a L\'opez, R. J., Randich, S. 2004a, A\&A, 418, 703  

\bibitem[2004b]{Ecu04b}
Ecuvillon, A., Israelian, G., Santos, N. C., Mayor, M., Villar, V., Bihain, G. 2004b, A\&A, 426, 619

\bibitem[2005a]{Ecu05a}
Ecuvillon, A., Israelian, G., Santos, N. C., Shchukina, N. G., Mayor, M., Rebolo, R. 2005a, A\&A, in press, astro-ph/0509326

\bibitem[2005b]{Ecu05b}
Ecuvillon, A., Israelian, G., Santos, N. C., Mayor, M., Gilli, G. 2005b, in Haute Provence Observatory Colloquium: Tenth Anniversary of 51Peg-b: status of and prospects for hot Jupiter studies, ed. L. Arnold, F. Bouchy and C. Moutou, in press, astro-ph/0509494

\bibitem[1993]{Edv93}
Edvardsson, B., Andersen, J., Gustafsson, B., Lambert, D. L., Nissen, P. E., Tomkin, J. 1993, A\&A, 275, 101

\bibitem[2004]{Egg04}
Eggenberger, A., Udry, S., \& Mayor, M. 2004, A\&A, 417, 353

\bibitem[1998]{Fel98}
Feltzing, S., \& Gustafsson, B. 1998, A\&AS, 129, 237

\bibitem[2005]{Fis05}
Fischer, D. A., \& Valenti, J. 2005, ApJ, 622, 1102

\bibitem[2005]{Gil05}
Gilli, G., Israelian, G., Ecuvillon, A., Santos, N. C., Mayor, M. 2005, A\&A, in press, astro-ph/0512219

\bibitem[1997]{Gon97}
Gonzalez, G. 1997, MNRAS, 285, 403

\bibitem[1998]{Gon98}
Gonzalez, G. 1998, A\&A, 334, 221

\bibitem[2001]{Gon01}
Gonzalez, G., Laws, C., Tyagi, S., \& Reddy, B. E. 2001, AJ, 121, 432

\bibitem[2003]{Gon03}
Gonzalez, G. 2003, Rev. Mod. Phys. , 75, 101

\bibitem[1999]{Gus99}
Gustafsson, B., Karlsson, T., Olsson, E., Edvardsson, B., \& Ryde, N. 1999, A\&A, 342, 426 

\bibitem[2001]{Isr01}
Israelian, G., Santos, N. C., Mayor, M., \& Rebolo, R. 2001, Nature, 411, 163

\bibitem[2003]{Isr03}
Israelian, G., Santos, N. C., Mayor, M., \& Rebolo, R. 2003, A\&A, 405, 753

\bibitem[2004a]{Isr04a}
Israelian, G. 2004a, in IAU S219: Stars as Suns: Activity, Evolution, and Planets, ed. A. K. Dupree (San Francisco: ASP), 343

\bibitem[2004b]{Isr04b}
Israelian, G., Santos, N. C., Mayor, M., \& Rebolo, R. 2004b, A\&A, 414, 601

\bibitem[2005]{Isr05}
Israelian, G. 2005, in Haute Provence Observatory Colloquium: Tenth Anniversary of 51Peg-b: status of and prospects for hot Jupiter studies, ed. L. Arnold, F. Bouchy and C. Moutou, in press

\bibitem[2001]{Jor01}
Jorissen, A., Mayor, M., \& Udry, S. 2001, A\&A, 379, 992

\bibitem[2000]{Lau00}
Laughlin, G. 2000, ApJ, 545, 1064

\bibitem[2001]{Law01}
Laws, C. \& Gonzalez, G. 2001, ApJ, 553, 405

\bibitem[2003]{Law03}
Laws, C., Gonzalez, G., Walker, K. M., Tyagi, S., Dodsworth, J., Snider, K. \& Suntzeff, N. 2003, AJ, 125, 2664

\bibitem[2003]{Lod03}
Lodders, K., 2003, ApJ, 591, 1220

\bibitem[1995]{May95}
Mayor, M. \& Queloz, D. 1995, Nature, 378, 355

\bibitem[2002]{Mur02}
Murray, N. \& Chaboyer, B. 2002, ApJ, 566, 442

\bibitem[2001]{Pin01}
Pinsonneault, M. H., DePoy, D. L., \& Coffee, M. 2001, ApJ, 556, 59

\bibitem[2000]{Que00}
Queloz, D., Mayor, M., Weber, L., Blecha, A., Burnet, M., Confino, B., Naef, D., Pepe, F., Santos, N.C., \& Udry, S. 2000, A\&A, 354, 99

\bibitem[2002]{Rei02}
Reid, I. N. 2002, PASP, 114, 306

\bibitem[2002]{Sad02}
Sadakane, K., Ohkubo, Y., Takeda, Y., Sato, B., Kambe, E., \& Aoki, W. 2002, PASJ, 54, 911

\bibitem[2002]{Sand02}
Sandquist, E. L., Dokter, J. J., Lin, D. N. C., \& Mardling, R. 2002, ApJ, 572, 1012

\bibitem[2000]{San00}
Santos, N. C., Israelian, G., \& Mayor, M. 2000, A\&A 363, 228

\bibitem[2001]{San01}
Santos, N. C., Israelian, G., \& Mayor, M. 2001, A\&A 373, 1019

\bibitem[2002]{San02}
Santos, N. C., Garc\'{\i}a L\'opez, R. J., Israelian, G., Mayor, M., Rebolo, R., Garc\'{\i}a-Gil, A., P\'erez de Taoro, M.
R., \& Randich, S. 2002, A\&A, 386, 1028

\bibitem[2003]{San03}
Santos, N. C., Israelian, G., Mayor, M., Rebolo, R., \& Udry, S. 2003, A\&A,398,363 

\bibitem[2004a]{San04a}
Santos, N. C., Mayor, M., Naef, D., Pepe, F., Queloz, D. \& Udry, S. 2004a, in IAU S219: Stars as Suns: Activity, Evolution, and Planets, ed. A. K. Dupree (San Francisco: ASP), 311

\bibitem[2004b]{San04b}
Santos, N. C., Israelian, G., \& Mayor, M. 2004b, A\&A, 415, 1153

\bibitem[2004c]{San04c}
Santos, N.C., Israelian, G., Garc\'{\i}a L\'opez, R. J., Mayor, M., Rebolo, R., Randich, S., Ecuvillon, A.,
\& Dom\'{\i}nguez Cerde\~na, C. 2004c, A\&A, 427, 1085

\bibitem[2005]{San05}
Santos, N. C., Israelian, G., Mayor, M., Bento, J. P., Almeida, P.C., Sousa, S. G., \& Ecuvillon, A. 2005,
A\&A, 437, 1127 

\bibitem[2001]{Smi01}
Smith, V. V., Cunha, K., \& Lazzaro, D., 2001, AJ, 121, 3207

\bibitem[2004]{Soz04}
Sozzetti, A. 2004, MNRAS, 354, 1194

\bibitem[2001]{Tak01}
Takeda, Y., Sato, B., Kambe, E., Aoki, W., Honda, S., Kawanomoto, S., \& Masuda, S., et al. 2001, PASJ, 53, 1211

\bibitem[2005]{Tak05}
Takeda, Y, \& Honda, S. 2005, A\&A, PASJ, 57, 65

\bibitem[2004]{Vau04}
Vauclair, S. 2004, ApJ, 605, 874

\bibitem[2003]{Udr03}
Udry, S., Mayor, M., \& Santos, N. C. 2003, A\&A, 407, 369

\bibitem[2002]{Zuc02}
Zucker, S., \& Mazeh, T. 2002, ApJ, 568, 113

\end{thebibliography}
\end{document}